# DeepCEST: 9.4 T Chemical Exchange Saturation Transfer MRI contrast predicted from 3T data - a proof of concept study

Moritz Zaiss[1], Anagha Deshmane[1], Mark Schuppert[1], Kai Herz[1], Philipp Ehses[2], Tobias Lindig[4], Benjamin Bender[4], Ulrike Ernemann[4], Klaus Scheffler[1,5]

[1]High-field Magnetic Resonance Center, Max Planck Institute for Biological Cybernetics, Tübingen, Germany.

[2]German Center for Neurodegenerative Diseases (DZNE), Bonn, Germany.

[3]Department of Neurosurgery, Eberhard-Karls-University Tübingen, Tübingen, Germany

[4]Department of Diagnostic and Interventional Neuroradiology, Eberhard-Karls University Tübingen, Tübingen, Germany.

[5]Department of Biomedical Magnetic Resonance, Eberhard-Karls University Tübingen, Tübingen, Germany.

Word count:   3500

*Corresponding Author

Moritz Zaiss

High-field Magnetic Resonance Center

Max Planck Institute for Biological Cybernetics

Tübingen, Germany

moritz.zaiss@tuebingen.mpg.de

+497071 601735

**List of abbreviations**

| | |
|---|---|
| ANN | artificial neural networks |
| CEST | chemical exchange saturation transfer |
| FLAIR | fluid attenuation inversion recovery |
| GRE | gradient-echo |
| MT | magnetization transfer |
| NOE | nuclear Overhauser effect |
| RMSE | root mean squared error |
| UHF | ultra-high field |

## Abstract

**Purpose:** Separation of different CEST signals in the Z-spectrum is a challenge especially at low field strengths where amide, amine, and NOE peaks coalesce with each other or with the water peak. The purpose of this work is to investigate if the information in 3T spectra can be extracted by a deep learning approach trained by 9.4T human brain target data.

**Methods:** Highly-spectrally-resolved Z-spectra from the same volunteer were acquired by 3D-snapshot CEST MRI at 3 T and 9.4 T with similar saturation schemes. The volume-registered 3 T Z-spectra-stack was then used as input data for a 3-layer deep neural network with the volume-registered 9.4 T fitted parameter stack as target data. The neural network was optimized and applied to training data, to unseen data from a different volunteer, and as well to a tumor patient data set.

**Results:** A useful neural net architecture could be found and verified in healthy volunteers. The principle gray-/white matter contrast of the different CEST effects was predicted with only small deviations. The 9.4 T prediction was less noisy compared to the directly measured CEST maps, however at the cost of slightly lower tissue contrast. Application to a tumor patient measured at 3 T and 9.4 T revealed that tumorous tissue Z-spectra and corresponding hyper/hypo-intensities of different CEST effects can also be predicted.

**Conclusion:** Deep learning might be a powerful tool for CEST data processing and deepCEST could bring the benefits and insights of the few ultra-high field sites to a broader clinical use. Vice versa deepCEST might help for determining which subjects are good candidates to measure additionally at UHF.

# Introduction

Chemical exchange saturation transfer (CEST) allows for indirect detection of solute molecules via exchanging protons that transfer selectively applied saturation to the large water pool in tissue. While studies have been performed at clinical field strengths, CEST effects were always also studied at higher field strengths where peak separation and selective saturation is easier (1–4). Insights from ultra-high field (UHF) scans helped to identify the origin of the detectable insights at lower fields, such as the detection of nuclear Overhauser effects (NOE) upfield form water (4,5), creatine and protein contributions at 2 ppm downfield form water (6–10), or glutamate and amine contributions around 3 ppm downfield from water (11). Thus, the knowledge gained at ultra-high fields always helped the interpretation of clinical CEST contrasts.

Some of the peaks can be detected separately only at ultra-high field and lead to the understanding that some signals detected at 3T are actually mixed signals from several resonances. The information is not gone, but just hard to extract, and 3T signals are still rich in information from different origins (12). The present article follows the approach of using prior UHF knowledge for 3T evaluation, in this case by applying artificial neural networks (ANNs) to combine these different data. The proposed neural network is trained using 3T Z-spectra as an input and 9.4T CEST parameters as a target. Thus, it is trained to predict 9.4T CEST contrasts from a Z-spectrum measured at 3T. In a way, this is the most direct approach of using the 9.4T prior knowledge for 3T data evaluation.

While application of neural networks in the field of MR has gained more and more interest in recent years (13–15), the presented approach represents a first step towards application in CEST MRI and is a rather simple approach. A multi-layer perceptron is used to combine co-registered data acquired at a 9.4T human scanner and a 3T clinical scanner. The prediction is tested in a second healthy subject and as well in a brain tumor patient. The proof of concept demonstrated herein not only hints to a future of UHF-guided evaluation tools applicable to clinical data, but also immediately provides decision support for the question of which patients measured at 3T might benefit from a UHF CEST scan. The presented "deepCEST" approach not only gives insight into what information is hidden in the acquired 3T data, but also widens the perspective of UHF centers as potential prior-knowledge generators for many clinical sites.

# Methods

Measurements were performed on 2 healthy subjects and on 1 patient with a brain tumor, with informed consent provided prior to the MRI experiments and under approval by the ethics committee of the medical faculty of the University Clinic, Tübingen Germany. All experiments were performed in accordance with relevant guidelines and regulations.

**9.4 T CEST MR imaging**

9.4 T CEST imaging was performed on a whole body MRI system (Siemens Healthcare, Erlangen, Germany). A custom-built head coil was used for signal transmission/reception (16 Tx/31 Rx channels) (16). The optimized 3D snapshot-CEST(17) acquisition consists of a pre-saturation module of 4.5 s followed by a readout module of duration $T_{RO}$ = 2.5 s, in which a train of RF and gradient spoiled gradient echoes with centric spiral reordering was acquired. Imaging parameters were FOV=220x180x32 mm$^3$, matrix size 144, 80% FOV in the first phase-encoding direction, thus, a phase-encoding matrix of 18x96 with phase encoding acceleration factor 3 and elliptical scanning; TE=1.85 ms, TR=3.64 ms, BW=700 Hz/px, 18 slices, FA=5° and elongation factor E=0.6 (rectangular spiral reordered).

The spectrally but not spatially selective CEST saturation period consists of a train of 150 Gaussian-shaped RF pulses, using a pulse duration of $t_{pulse}$ = 15 ms separated by pulse delay of $t_{delay}$ = 15 ms, resulting in a total saturation time of $T_{sat}$ = 4.5 s, and nominal $B_1$ values of $B_1$ = 0.6 µT, 0.9 µT, 1.2 µT. After the pulse train, a crusher gradient was applied to spoil residual transversal magnetization. Z-spectrum data was obtained after saturation at 95 offsets in the range of ±50 ppm with denser sampling in the range of ±5 ppm (17) and normalized by unsaturated scans with 12s of relaxation and saturation at -300 ppm. After each acquisition a recovery time of $T_{rec}$ = 1.1 s takes place. Acquisition time per offset was TA= $T_{rec}$ +$T_{sat}$+$T_{RO}$=8.1 s. For 95 offsets this yields a total scan time of approximately $TA_{tot}$ = 12 minutes for the total high-resolution CEST Z-spectrum scan. For CEST at three different $B_1$ levels, plus $B_0$ and $B_1$ mapping using WASABI (18) and $T_1$ mapping using a saturation recovery sequence, the total examination time of the highly-resolved CEST protocol was 40 minutes. During the patient scan, the CEST images were acquired at only two $B_1$ levels due to limited scan time ($B_1$ = 0.9 µT, and 1.2 µT).

**3 T CEST MR imaging**

CEST imaging at 3 T was performed on a whole body MRI system (PRISMA, Siemens Healthcare, Erlangen, Germany) on the same 2 healthy subjects and one patient. A similar 3D snapshot-CEST acquisition optimized for 3T consisted of a pre-saturation module of 4 s followed by a readout module of duration $T_{RO}$ = 3.5 s, in which a train of RF and gradient spoiled gradient echoes with centric spiral reordering was acquired. Imaging parameters were FOV=220x180x48 mm$^3$, matrix size 128, 80% FOV in the first phase-encoding direction, phase encoding acceleration factor 2 and elliptical scanning; TE=2 ms, TR=5 ms, BW=400 Hz/px, 18 slices, FA=6° and elongation factor E=0.5 (rectangular spiral reordered).

The spectrally but not spatially selective CEST saturation period consists of a train of 80 Gaussian-shaped RF pulses, using a pulse duration of $t_{pulse}$ = 20 ms separated by pulse delay of $t_{delay}$ = 20 ms, resulting in a total saturation time of $T_{sat}$ = 3.6 s, and a single nominal $B_1$ value of $B_1$ = 0.6 µT. After the pulse train, a crusher gradient was applied to spoil residual transversal magnetization. Z-spectrum data was obtained after saturation at 56 offsets in the range of ±100 ppm with denser sampling in the range of ±5 ppm (17) and normalized by unsaturated scans with 12s of relaxation and saturation at -300 ppm. With $B_0$ and $B_1$ mapping using WASABI (18) and $T_1$ mapping using a saturation recovery sequence, the total examination time for CEST at 3T was 10.5 minutes.

**Data evaluation**

Z-spectrum data was first corrected for motion using the AFNI's 3Dvolreg function (19), followed by $B_0$ and $B_1$ inhomogeneity correction using the WASABI approach (18), and the Z-3-point-$B_1$-correction method (20) (images reconstructed at 0.5 µT for 9.4T). Reference images are then manually masked to isolate CSF and brain tissues. CEST images were generated from the Z-value $Z(\Delta\omega)$, given by the ratio of the saturated image $S_{sat}(\Delta\omega)$ and the unsaturated image $S_0$

$$Z(\Delta\omega) = \frac{S_{sat}(\Delta\omega)}{S_0} \qquad (1)$$

To isolate CEST effects in the 9.4T Z-spectra from direct water saturation (spillover) and semi-solid magnetization transfer (ssMT), the five pool Lorentzian fitting method was used according to (21):

$$Z(\Delta\omega) = c - L_w - L_{mt} - L_{2.0} - L_{-1.6} - L_{-3.5} \qquad (2)$$

with each $L_x$ being a Lorentzian function defined by

$$L_x = A_x \frac{\Gamma_x^2/4}{\Gamma_x^2/4 + (\Delta\omega - \delta_x)^2} \qquad (3)$$

with the amplitude $A_x$, the full-width at half maximum $\Gamma_x$, and the chemical shift $\delta_x$ of the proton pool x relative to the water proton pool. Fit starting values and boundaries for each Lorentzian of this six-pool fit are given in the Supporting information including a description of how they were determined. Maps of each CEST peak refer then to the individual Lorentzian amplitude $A_x$ in each pixel. Fitting was performed using least-squares optimization in MATLAB (R2016a, The Mathworks Inc, Natick, MA, USA) and took about 5 mins per 3D stack of Z-spectra.

**Neural networks architecture and training**

The principal artificial neural network architecture is a multi-layer perceptron as shown in Figure 1. We define the *deepCEST network* as a network where the *input data* is a 3T Z-spectrum and the *output* is a vector of Lorentzian fit parameters from Equation 3 obtained by an ultra-high field Z-spectrum fit. In Figure 1, each node represents a neuron, and each connection between the nodes represents a so-called layer weight. In a fully connected network as displayed here, the layer weights for each layer form a matrix $LW_{ij}$. The parameters in these matrices $LW_{ij}$ are the free parameters of the neural network that are adjusted during the training process. In the base implementation, the *deepCEST* network takes a 3T Z-spectrum with 45 offsets as input, yields 19 Lorentzian fit parameters as output, and has 3 layers and 400 neurons total, resulting in 54·100+100·200+200·100+100·19 = 47300 free parameters.

For a given input vector **Z**, the activations of the hidden layer Neurons are given by the following sigmoidal relations that are subsequently evaluated.

$$\mathbf{A1} = \mathrm{tansig}(\mathbf{b1} + \mathbf{ILW} \cdot \mathbf{Z}) \tag{4}$$

$$\mathbf{A2} = \mathrm{tansig}(\mathbf{b2} + \mathbf{LW1} \cdot \mathbf{A1}) \tag{5}$$

$$\mathbf{A3} = \mathrm{tansig}(\mathbf{b3} + \mathbf{LW2} \cdot \mathbf{A2}) \tag{6}$$

Where **ILW** are the initial layer weights, **LW1, LW2** are hidden layer weights, **b1, b2, b3** are additional bias neurons (bias neurons are not shown in Figure 1), and *tansig* is the activation function given by

$$\mathbf{tansig(x)} = 2/(1+\exp(-2x)) - 1 \tag{7}$$

The output is then given by the activations **A3** and the so-called regression layer

$$\mathbf{out} = \mathrm{mapminmax}(\mathbf{b4} + \mathbf{LW3} \cdot \mathbf{A3}) \tag{8}$$

where the function *mapminmax* is a linear function matching the interval [-1 1] to the range of the ouput data.

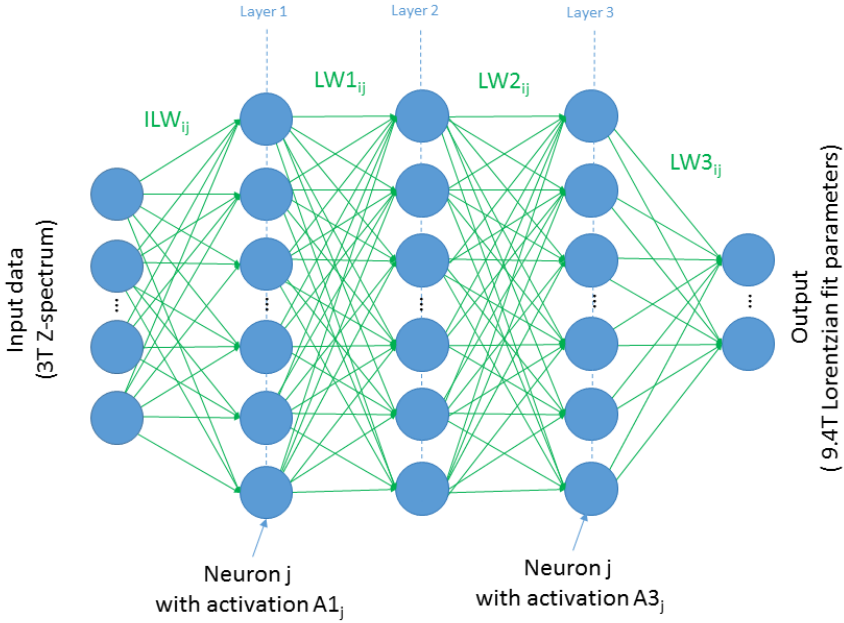

**Figure 1:** Scheme of the *deepCEST* network with an input 3T Z-spectrum (54 nodes) and output vector 19 9.4T-Lorentzian-fit parameters in Equation 3. This network consists of 3 layers with 100, 200, and 100 neurons, respectively. The free parameters of the network are illustrated by the green arrows that represent the layer weights. The *deepCEST* network with 3 layers and 400 neurons has 54·100+100·200+200·100+100·19 = 47300 free parameters.

The training process is a so-called back-propagation optimization (scaled conjugate gradient backpropagation (22)) of the free parameters of the network using input 3T Z-spectra and target 9.4T parameter vectors of a training dataset. The starting point of the training uses randomly initialized matrices. To avoid overfitting, two strategies were employed. The first is *early stopping*. Here, the training data is randomly divided into a training set (70%), a validation set (15%) and a test set (15%), and the validation set is used to determine an early stopping criteria. If the root mean square error (rmse) of the validation data does not improve within 5 iterations, the optimization is stopped. The second method is *regularization*, which uses a regularization factor $\gamma$ to add a penalty to large weights and thus avoid overfitting. If *mse* is the mean-squared-error of the optimization and *msw* are the mean-squared-weights then the optimized function with regularization is *msereg=γ∗msw+(1−γ)∗mse*. In the final deepCEST network training, $\gamma$ was set to 0.5.

Input data of the *deepCEST* neural network are 3T Z-spectra of 54 frequency offsets, and the output or target data were the 19 Lorentzian fitting parameters of a 9.4T scan with 97 frequency offsets. Several neural networks were tested and are reported in the Supporting Information. The number of layers as well as the number of neurons in each layer was varied and optimized as well as the regularization factor $\gamma$ in the training procedure (see Supporting Figure S1). The final architecture was chosen as a compromise between good performance and fast learning, leading to a regularization factor of $\gamma$ =0.5 and a 3-layer-network with neurons in each layer given by

[100 200 100] where this notation stands for [(neurons in input layer) (neurons in hidden layer) (neurons in output layer)].

For training, data from only one volunteer was used. The second volunteer's data as well as the tumor patient data were reserved for an independent test of the network.

## Results & Discussion

Visualization of the trained neural network in action with layer weights and activations as given in Equations 4-8 is given in Figure 2. Please note, the layer weights are adjusted during training but constant for the trained net, and the activations depend on the specific input. Here we show the structure and parameters of the trained neural network together with its application to a 3T input Z-spectrum. The constant initial layer weights (Figure 2b) can be understood as filters applied to the Z-spectra to extract certain features. This initial layer shows some symmetric and asymmetric filters visibly highlighting features at direct saturation/semisolid-MT measurement indices. The hidden layer weights (Figure 2d,f) show the recombination of the initial gathered features. The output layer weights show that for each parameter, several features are used (Figure 2h). The output layer activations lead then to the output parameters employing a linear function (Eq. 9) displayed in Figure 2j. With these parameters and the Lorentzian function (Eq. 2-3) a predicted Z-spectrum can be calculated (solid red line, Figure 2a).

After optimization of this neural network architecture (see Supporting Figure S1) a final network was trained and the training data set was analyzed (Figure 3). While predicted and real Z-spectra in three ROIs show a visual match (Figure 3 a-c), the prediction from the training dataset still shows small differences from the used target dataset, which are better visible in the parameter maps (Figure 3 d-g, h-k). Most obvious is the denoised appearance of the predicted data (Figure 3 h-k). For more quantitative insight, the difference between real and predicted contrasts was calculated (Figure 3, l-o). This reveals systematic changes e.g. in the CSF regions and also visible in the ROI evaluations (Figure 3, a-c).

To analyze the dependencies of prediction mismatch of the neural network training, the exact same network architecture was trained with the same training data 10 times with different random starting weights. Figure 4 shows application of the 10 nets to the same data as in Figure 3. ROI evaluations show that variation between different nets is small in tissue, with stronger deviations observed in CSF, also demonstrated by the higher standard deviation especially in CSF regions (Fig 4 q,r). The ssMT prediction shows to have the strongest standard deviations across networks. Still, the overall standard deviations of the training process are all below 1% and thus smaller than the deviation from the data. This means that the variability of the training process is much smaller than the variability of the output of a single trained net over the image. The mean prediction shows similar outcome as the single net in Figure 3. From this analysis, we can conclude that there are nets with better and worse predictions, and that CSF pixels can cause problems. An explanation for the variations of the trained nets in CSF could be the lower amount of CSF data, as well as pulsations and movement that are expected to be more severe in the CSF regions. The stronger deviations of the ssMT could be an interplay between the broad semi-solid line and the additional baseline constant used in the fit. However, observed differences also incorporate co-registration mismatches of the 3T and 9.4T data and represent an upper limit of the network's prediction errors. For the following evaluations, the mean network could be used, but to save computation time we decided for the network reflecting the closest match to the mean network outcome.

In a next step, the trained network was applied to a completely new 3T dataset of a second volunteer to verify its generalization. Figure 5 shows the predicted data together with the additionally acquired 9.4T data. The measured data shows some regional artifacts that resulted from residual motion and B0 artifacts that could not be corrected for. Predicted amide and NOE contrast show the expected grey white matter contrast, but especially the amine CEST prediction shows reduced GM/WM contrast compared to the measured data and the training data set (Figure 3). CSF regions show deviations similar to those shown in Figure 3. Z-spectra are astonishingly similar in GM and WM tissues, but again the CSF ROI prediction as well as the overall ssMT estimation shows the largest deviations.

A pixel wise-correlation of the whole test data (details summarized in Figure 8) set revealed strong correlation of the *deepCEST* prediction with the actual 9.4T data ($R^2=0.88$) close to the $R^2$ of the training data of 0.95. Thus, we can conclude that the net trained on the data of a single volunteer not only performs well on training data, but is able to generally predict 9.4T contrast from 3T data with some loss in contrast. At 9.4T, $B_0$ and $B_1$ inhomogeneities are quite problematic and require correction that is already incorporated in the 9.4T data. Notably, the 9.4T maps predicted from 3T data have the nice $B_1$ and $B_0$ homogeneity features of the 3T scan.

As an ultimate test for generalization of the *deepCEST* prediction ability, a tumor patient was scanned at both field strengths and the net, trained only on data from the first healthy subject, was applied to it. Figure 6 shows that the UHF CEST prediction in the tumor patient is comparable to the healthy test volunteer shown in Figure 5. In addition to healthy tissue contrast, tumor tissue contrast can also be predicted by the neural network: amide CEST shows hyper-intensity in the tumor area in both acquired 9.4T data as well as in the deepCEST prediction. Also, NOE-CEST and ssMT maps show clear decrease in the tumor area as well as a strong drop in the necrotic cyst as expected (21,23). At the boundaries of the cyst, the real data shows some highlights also observed in the prediction; these features are better visible in the zoomed image in Figure 7 a,b (black arrows). Figure 7c,d show the 2 ppm signal with a more narrow windowing, revealing similar hyperintensities in the tumor area when compared to contralateral tissue. However, some small regions are wrongly predicted (white arrows). It seems that CSF-like tissue is predicted as this region is also hypointense in the NOE prediction. The 2 ppm signal might still be least reliably extracted from 3T data, as the resonance is strongly affected by direct water saturation at 3T. The regression analysis shows now even lower $R^2$ of 0.83 and the predicted features are generally lower than those observed in the measured 9.4T data (see Figure 8c).

The approach presented herein is not employing spatial information, but single voxel Z-spectral data. Thus the achieved spatial coherence in tissues in the test data sets is already an interesting result and a verification of the generality of the *deepCEST* nets. Also, in contrast to neural nets working on images that demand many brains and tumors to come into the big data regime, the voxel-frequency-based approach already has 200,000 input and target vectors in the 3D training data set of a single volunteer available for training. Moreover, while image based neural network approaches need to be trained with tumor data, we could show that a Z-spectrum based neural network trained with healthy data was also able to predict tumor Z-spectra and yield enhanced or depleted contrast similar to the real 9.4T scan. One could ask how the neural network can predict signals that it has never learned; however, this is equivalent to the question if tumor Z-spectra can be expressed as superposition of Z-spectra of different healthy tissues, which is more plausible. Thus, already with one or a few volunteer training data sets, a relatively general *deepCEST* network can be created. Of course, some patient datasets could be added to the training. However, we think it is beneficial to not do so: if patient data is added it can be argued that the net predicts tumors because they were in the training and the result is therefore biased. A neural network that predicts contrast in tumors but is only trained on healthy volunteers is therefore favorable.

While the neural network functionality is not very transparent, the *average gradient method* as described in the Supporting Information can be used to gain insights into which frequency offsets the neural network is more sensitive to for each parameter determination. The average gradient of the network is shown in Supporting Figure S2 and shows at least plausible offset sensitivities for both NOE and amide CEST effects with coarse correlation to the GM/WM difference Z-spectra. If the network would have disproportionately weighted a single offset to estimate the output parameters, such an offset would be visible in Figure S2. In principle, it is possible to further analyze the network to optimize the actual acquired data points. Thus, the *deepCEST* network is not only a useful tool for generating contrast, but also allows further insights into where the information is hidden at 3T.

The *deepCEST* approach presented here is configured for low power Z-spectra that are dominated by protein signals. However, the principle approach can be translated to creatine, glutamate, or hydroxyl CEST measurements which were until now also most successful at UHF (6,7,11,24,25). Finally, nothing besides an actual UHF scan can give the insight of an UHF scan. However, while the predicted contrasts cannot be used yet for clinical decision making, it could already be of use for the imaging diagnostician: if the predicted UHF *deepCEST* contrast shows interesting features, this information could be used to plan an actual UHF CEST scan for this patient.

# Conclusion

Prior knowledge about CEST spectra gained at UHF can be elegantly translated to 3T scans by using a pretrained *deepCEST* neural network to extract CEST contrasts. Neural nets need to be complex enough to describe multi-pool Z-spectra, while reducing complexity by reducing neurons enables denoising features. Using spectral data from each voxel and training with one co-registered 3D data set from a healthy volunteer, the net already allows for sufficient generality to predict unseen subject data and even tumor Z-spectra. Thus, the proof of concept is given; still, some artifacts could be observed in the prediction, thus the approach must be carefully further optimized. While clinical application of this approach might be validated in the future, the *deepCEST* prediction can already help in the decision of which patient could benefit from an additional UHF CEST scan.

## Acknowledgments

The financial support of the Max Planck Society, the German Research Foundation (DFG, grant ZA 814/2-1, support to M.S., K.H.), and the European Union's Horizon 2020 research and innovation programme (Grant Agreement No. 667510, support to M.Z., A.D.) is gratefully acknowledged.

# List of Figures

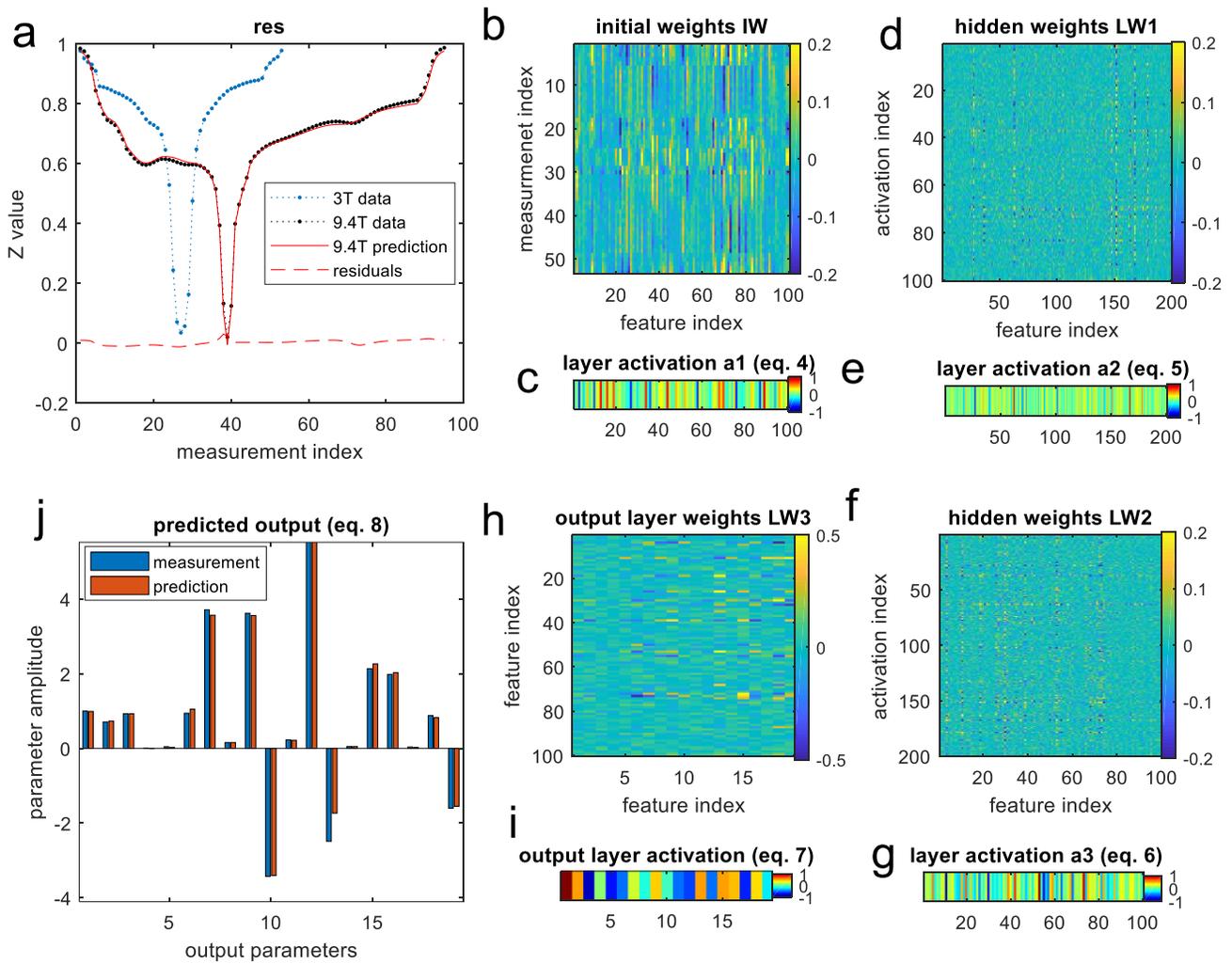

**Figure 2:** Visualization of the trained *deepCEST* neural network application as given by Equations 4-8, with layer weights and neuron activations as given in Figure 1. The 3T input data (blue line (a)) is feed into the first layer of the neural net (b); please note that the x-axis corresponds to measurement number and not to the offset frequency. The layer weights of the initial layer (b) show the filters which the training process generated to filter the 3T-Z-spectra, such that the specific 3T Z-spectrum in (a) yields the activation patter in (c) in the 100 neurons of the initial layer. These neurons are sigmoidally connected to the second layer with 200 neurons with (d) showing the layer weights for each connection between the first and second layers. With the given input this yields to the activation of layer 2 shown in (e). Layer weights and activation of layer 3 are given in (f) and (g). Layer 3 is connected to the output layer (h) with activations given in (i) which is linearly connected (Eq. 9) to the output parameters for this input, shown in (j). With these output parameters the predicted Z-spectra (solid line in a) was calculated and is compared to the actual measured 9.4T Z-spectrum (black points in a).

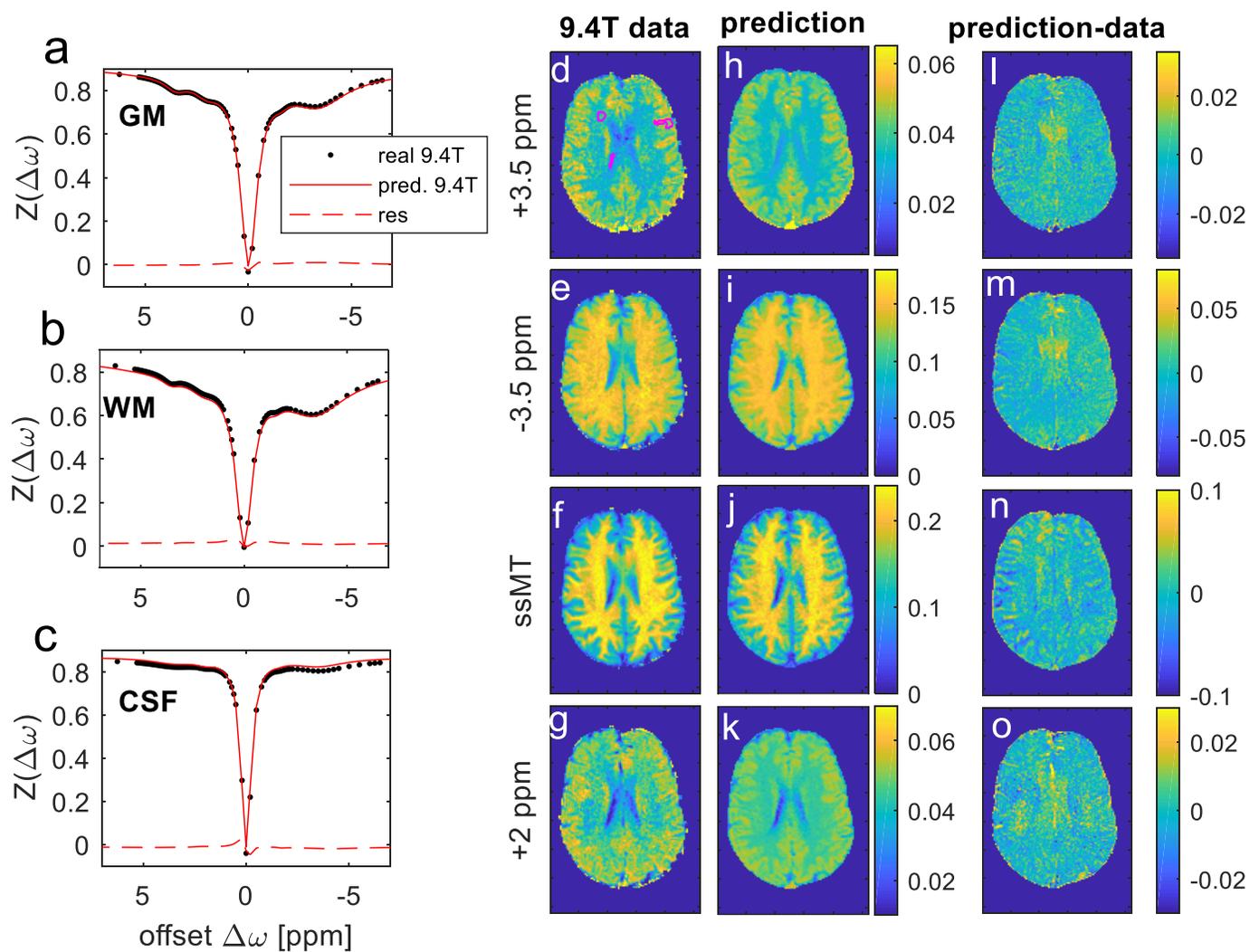

**Figure 3:** *deepCEST* network applied to one healthy subject training dataset. ROI evaluations (a-c), real 9.4T CEST fits (d-g), prediction of net (h-k), prediction difference from data (l-o),

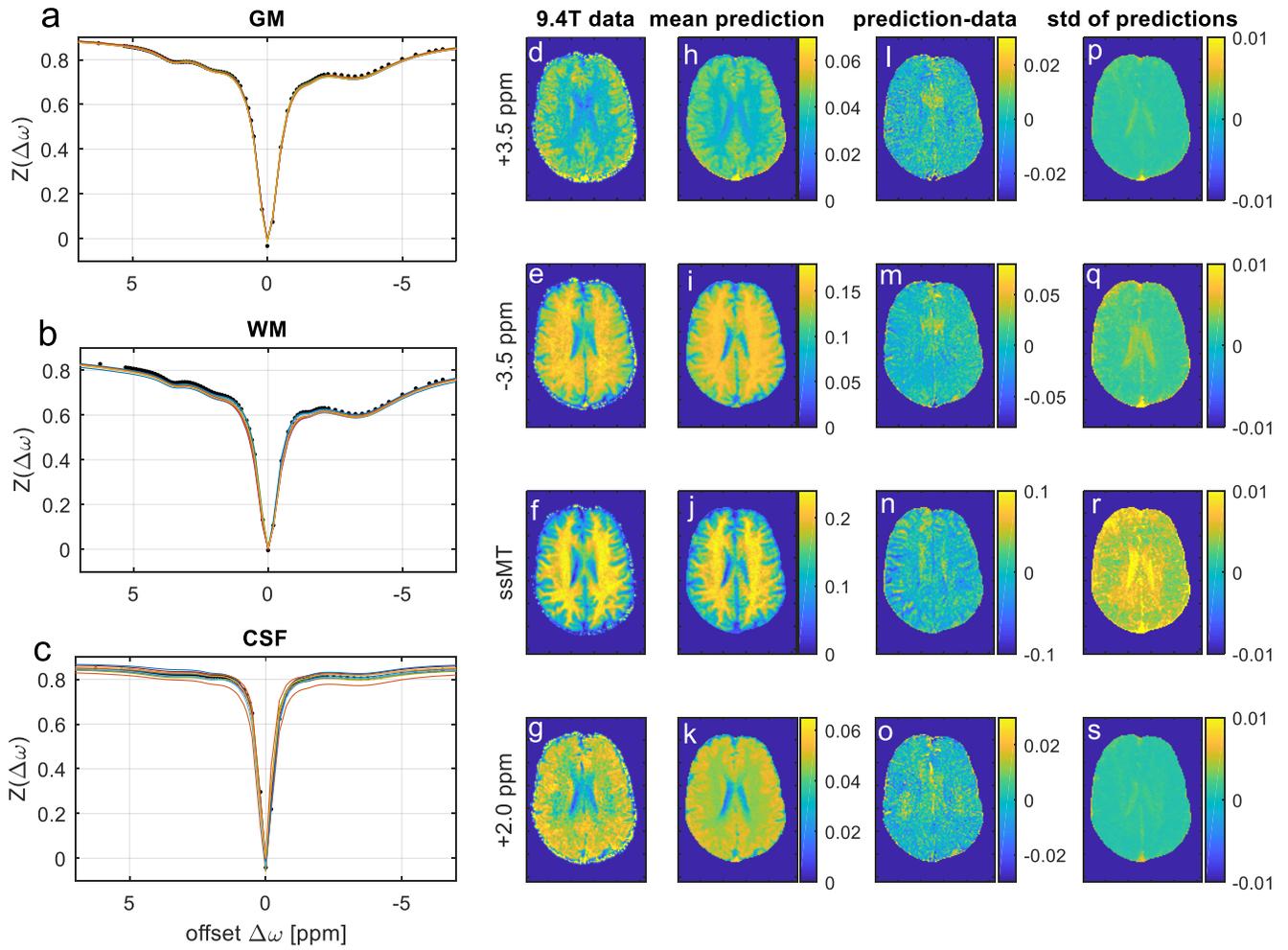

**Figure 4:** Variation of network training analyzed using 10 trained nets applied to 3T training data. ROI evaluations (column 1, a-c), real 9.4T data (column 2, d-g), mean prediction of the 10 nets (column 3, h-k), mean prediction difference form data (column 4 l-o), and the standard deviation of the 10 predictions (p-s).

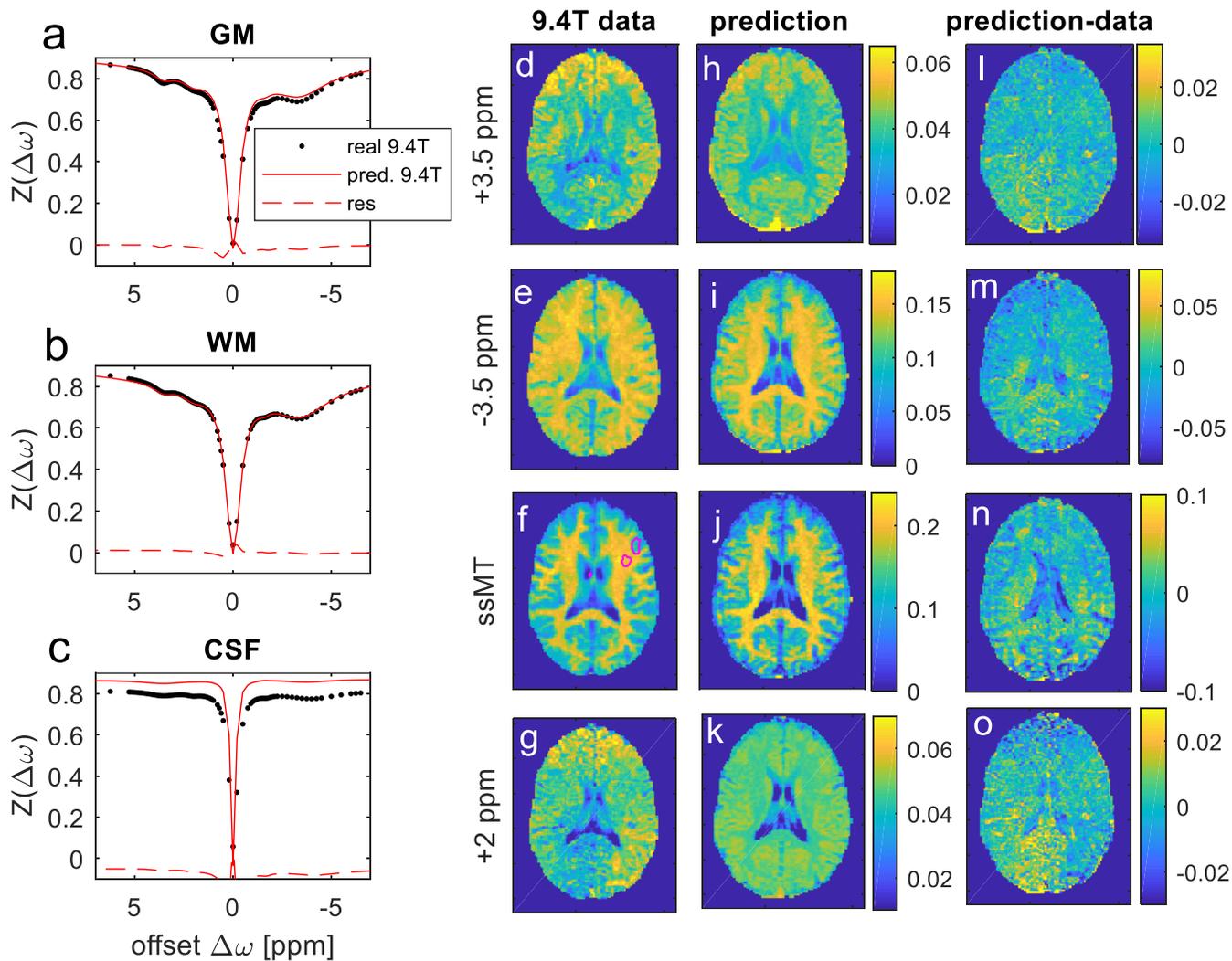

**Figure 5:** *deepCEST* application to test data from a second healthy subject. ROI evaluations (a-c), real 9.4 T data (d-g), prediction of net (h-k), prediction difference from data (l-o).

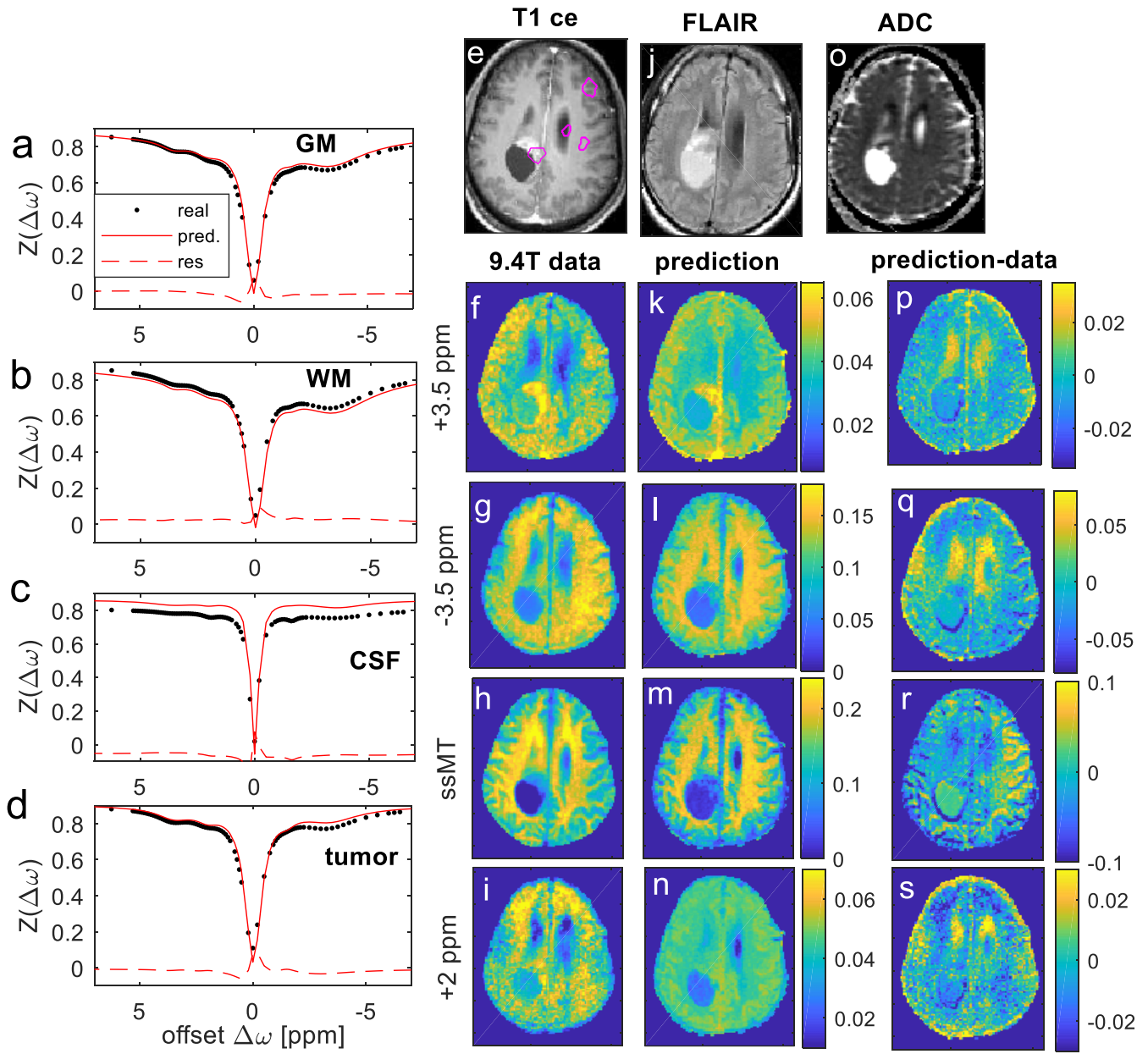

**Figure 6:** deepCEST application to subject with brain tumor test data. ROI evaluations (a-d), clinical contrasts (e,j,o), real 9.4T data (f-i), prediction of net (k-n), prediction difference from data (p-s).

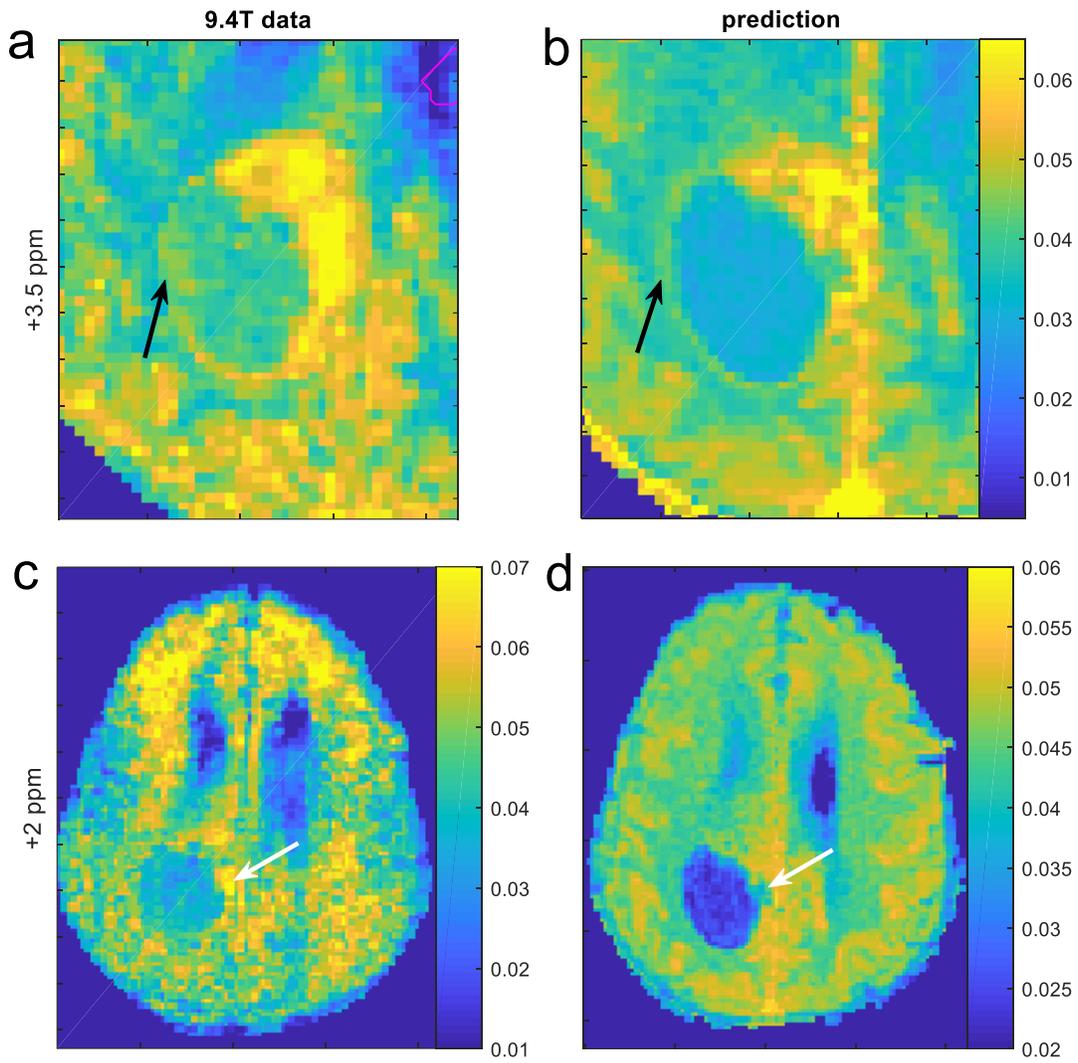

**Figure 7:** (a,b) Zoomed version of subfigures (f) and (k) of Figure 6. (c,d) tighter windowed versions of figures (i) and (n) of Figure 6. Black arrows show the enhancement at the edge of the cyst visible in both prediction and real data. White arrows show an area that is predicted hypo intense, but measured hyper intense in the 2 ppm signal.

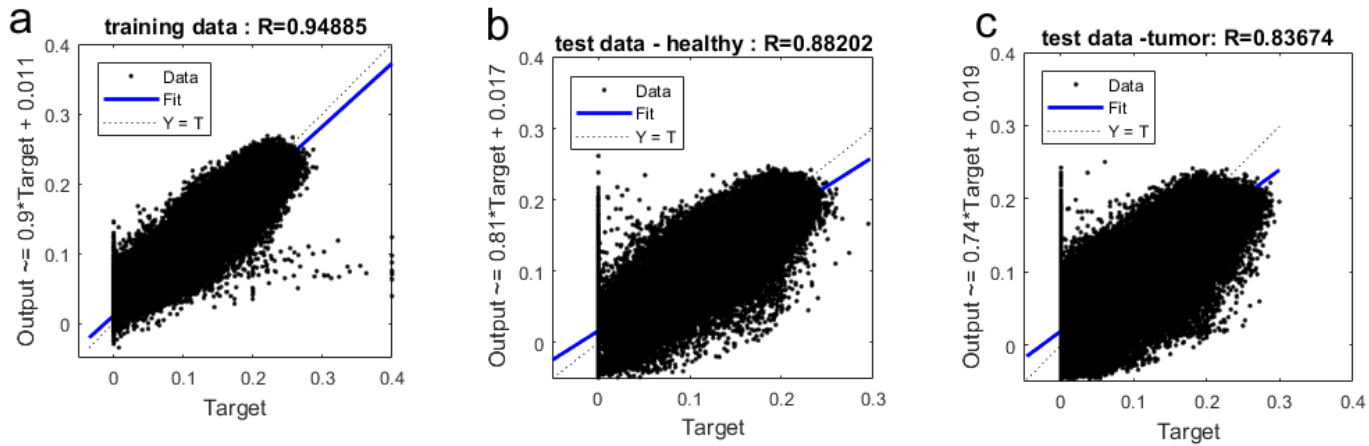

**Figure 8:** Regression plots for amide, NOE and ssMT signals for training data (a), test data of healthy subject (b), test data of subject with brain tumor (c).

## Supporting information

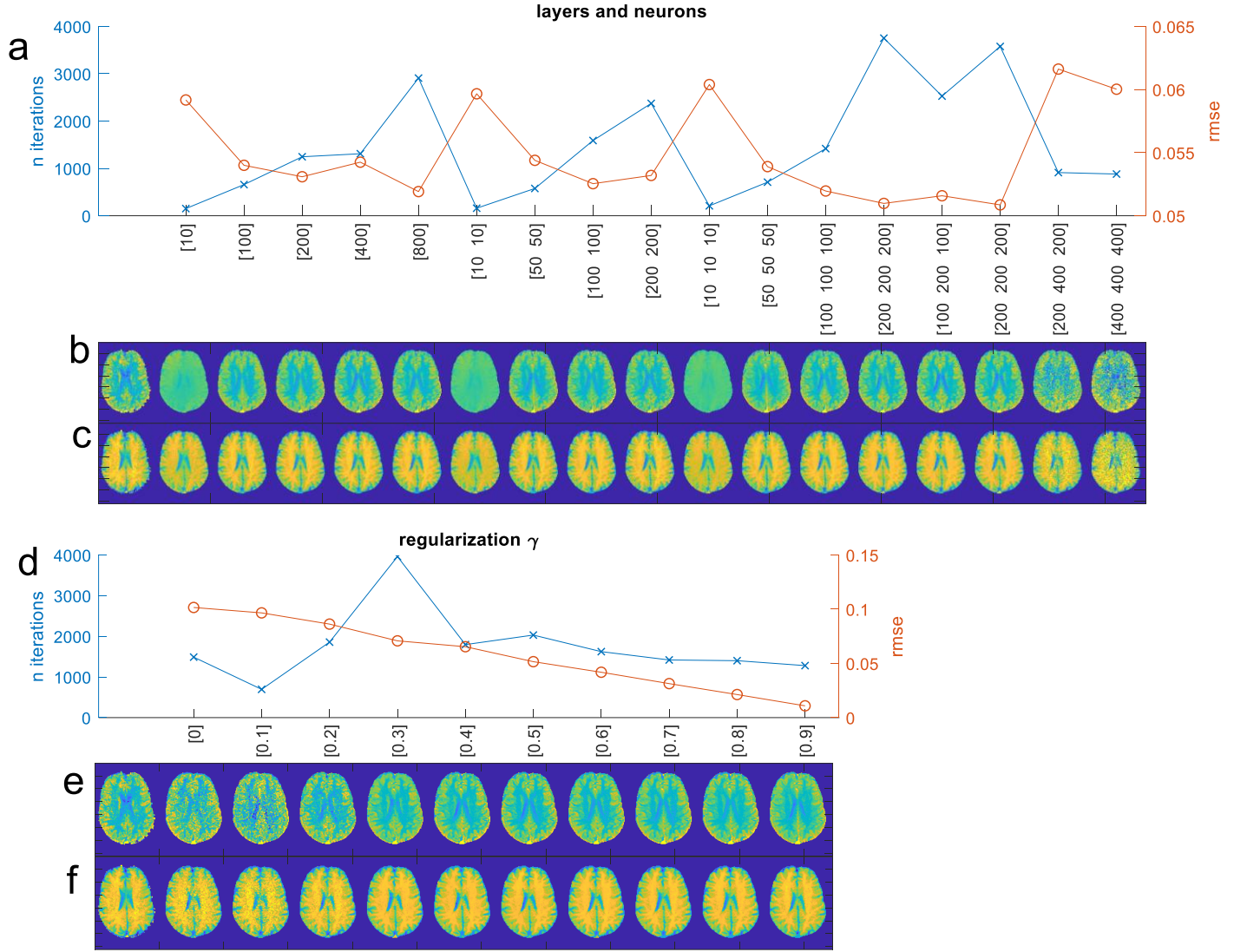

**Figure S1:** Performance (rmse) and number of iterations during training of different network architectures (a) and regularization fixed at γ=0.5, as well as for different regularization terms for the network [100 200 100] (d). Panels (b,c) and (e,f) show the corresponding prediction of the amide CEST(b,e) and NOE CEST contrast (c,f) together with the real data (leftmost images).

Optimization of the neural network architecture (Figure S1 a-c), and regularization (Figure S1 d-f) shows that 3 layers yield the lowest root-mean-squared-error (rmse). The regularization yields good results for γ>=0.3. Thus, we choose 0.5 as a compromise for stronger penalty for high weights to achieve better generality. Both more neurons and lower regularization lead to an effectively higher number of free parameters which then allow the network to fit noise patterns in the training data, which can be observed as the noise in the predicted amide-CEST and NOE-CEST maps (figure S1 b,c and e,f). Thus, by limiting the effective number of parameters, denoising and generalization can be achieved. The used regularization factor of 0.5 and layer structure with neurons given by [100 200 100] seems to be a good compromise between performance, generalization, and denoising.

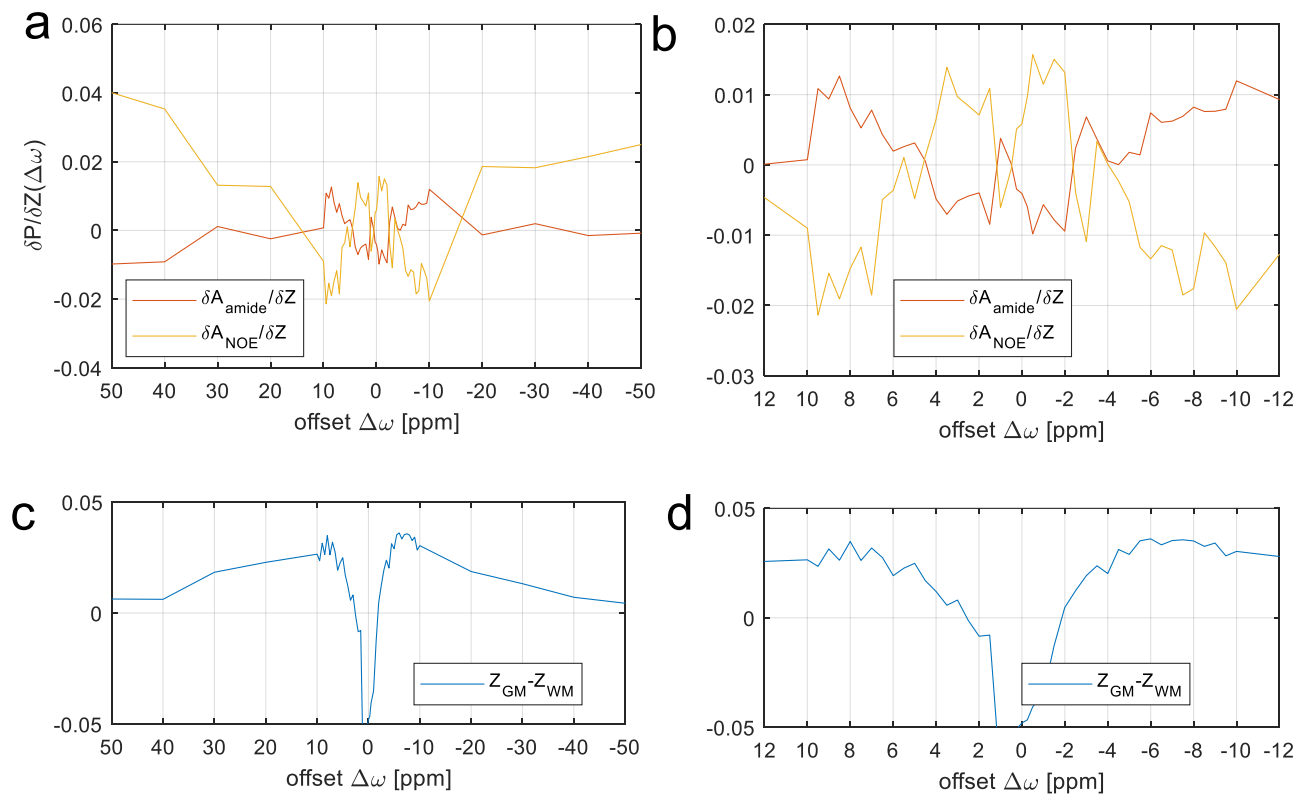

**Figure S2:** Network analysis by average gradient $\delta P/\delta Z(\Delta\omega)$ for the trained deepCEST network and N=M=40. Displayed are the amplitudes of the amide CEST and the NOE CEST peaks(a,b). Difference spectrum of grey and white matter (c,d). (a,c) zoomed out spectra, (b,d) zoomed in the range of -12 to 12 ppm.

As a neural network is highly non-linear, a simple gradient calculation is not accurate according to (ftp://ftp.sas.com/pub/neural/importance.html). Thus $\delta P/\delta Z(\Delta\omega)$, where P is the output parameter and Z is the input Z spectrum value, is estimated by evaluating many random input data with alterations of each individual offset. Pseudocode of this calculation is given below. $\delta P/\delta Z(\Delta\omega)$ is a matrix of size (number of output parameters × number of input offsets) and yields the average sensitivity of each parameter as a function of each offset Z-value:

### Pseudo-code for calculation of average gradient

```
Z0 = list of all training data Z-spectra;
nn=1;

for kk=1:N
    Z1=Z0(:,random_index);          % get N random 3T-Z-spectra from the dataset
    PZ1=net(Z1);                    % get deepCEST prediction for this spectrum

    for jj=1:M                      % get M random variations of Z1
        for ii=1:number_offsets  % for each offset

            Z2=Z1;                  % X2 refreshed in each iteration
            Z2(ii)=Z2(ii)+ 0.1*randn; % Z2(dw_ii) varied normally distrib.

            PZ2=net(Z2);            % get deepCEST prediction for altered spectrum

            dPdZ(:,ii,nn)=(PZ2-PZ1)./(Z2-Z1);  % calculate difference quotient

        end
        nn=nn+1;   % nn sums ii and jj, iteration variable for different Z1 and varied Z1=Z2
    end
end

δP/δZ(Δω) = mean(dPdZ (:,:,nn),nn); %average_gradient over all N and M variations
```

The average gradient for the used *deepCEST* network and N=M=40 is displayed in Figure S2 for the amide signal and the NOE signal. Both gradients for NOE and amide amplitude show dependencies for all offsets of the Z-spectrum. This is not surprising as a multi-Lorentzian fit also requires knowledge of the background signals to isolate selective effects. The overall shape of the gradients coarsely correlates (amide) or anti correlates (NOE) with the difference spectrum of grey and white matter (Figure S2c,d). The gradient $\delta A_{amide}/\delta Z(3.5ppm)$ is negative, meaning lower Z-values here lead to higher amide CEST effect and is thus plausible. The gradient $\delta A_{NOE}/\delta Z(-3.5ppm)$ does not show strong dependence on the input and NOE is more determined by offsets between -6 ppm and -10 ppm. This behavior can be understood by looking at the difference spectrum of grey and white matter, which also shows stronger differences around -6 to -10 ppm compared to -3.5 ppm. At 0 ppm the amide gradient has a sign change from negative to positive, and the NOE gradient has a sign change from positive to negative. This can be understood as a $B_0$ correction feature: the response of a slightly shifted direct water saturation line will average out. To summarize, at least some insight into the inner workings of the neural network can be gained, and plausible sensitivity patterns are also observed. More detailed analysis of such networks might help to identify the most important frequency offsets that should be measured.

# References

(1) Zhou J, Payen J-F, Wilson DA, Traystman RJ, Zijl PCM van. Using the amide proton signals of intracellular proteins and peptides to detect pH effects in MRI. Nat Med 2003;9:1085–1090.

(2) Zhou J, Lal B, Wilson DA, Laterra J, van Zijl PCM. Amide Proton Transfer (APT) Contrast for Imaging of Brain Tumors. Magn. Reson. Med. 2003;50:1120–1126.

(3) Jones CK, Polders D, Hua J, Zhu H, Hoogduin HJ, Zhou J, Luijten P, van Zijl PC. In vivo three-dimensional whole-brain pulsed steady-state chemical exchange saturation transfer at 7 T. Magnetic Resonance in Medicine 2012;67:1579–89.

(4) Jones CK, Huang A, Xu J, Edden RAE, Schär M, Hua J, Oskolkov N, Zacà D, Zhou J, McMahon MT, Pillai JJ, van Zijl PCM. Nuclear Overhauser Enhancement (NOE) Imaging in the Human Brain at 7T. Neuroimage 2013;77:114–124.

(5) Zaiss M, Kunz P, Goerke S, Radbruch A, Bachert P. MR Imaging of Protein Folding in Vitro Employing Nuclear-Overhauser-Mediated Saturation Transfer. NMR Biomed 2013;26:1815–1822.

(6) Kogan F, Haris M, Singh A, Cai K, Debrosse C, Nanga RPR, Hariharan H, Reddy R. Method for High-Resolution Imaging of Creatine in Vivo Using Chemical Exchange Saturation Transfer. Magnetic Resonance in Medicine 2014;71:164–172.

(7) Rerich E, Zaiss M, Korzowski A, Ladd ME, Bachert P. Relaxation-Compensated CEST-MRI at 7 T for Mapping of Creatine Content and PH--Preliminary Application in Human Muscle Tissue in Vivo. NMR in Biomedicine 2015;28:1402–1412.

(8) Cai K, Singh A, Poptani H, Li W, Yang S, Lu Y, Hariharan H, Zhou XJ, Reddy R. CEST Signal at 2ppm (CEST@2ppm) from Z-Spectral Fitting Correlates with Creatine Distribution in Brain Tumor. NMR Biomed 2015;28:1–8.

(9) Chen L, Zeng H, Xu X, Yadav NN, Cai S, Puts NA, Barker PB, Li T, Weiss RG, van Zijl PCM, Xu J. Investigation of the Contribution of Total Creatine to the CEST Z-Spectrum of Brain Using a Knockout Mouse Model. NMR Biomed 2017;30,DOI10.1002/nbm.3834.

(10) Zhang X-Y, Xie J, Wang F, Lin EC, Xu J, Gochberg DF, Gore JC, Zu Z. Assignment of the Molecular Origins of CEST Signals at 2 Ppm in Rat Brain. Magnetic Resonance in Medicine 2017;78:881–887.

(11) Cai K, Haris M, Singh A, Kogan F, Greenberg JH, Hariharan H, Detre JA, Reddy R. Magnetic Resonance Imaging of Glutamate. Nature Medicine 2012;18:302–306.

(12) van Zijl PCM, Lam WW, Xu J, Knutsson L, Stanisz GJ. Magnetization Transfer Contrast and Chemical Exchange Saturation Transfer MRI. Features and Analysis of the Field-Dependent Saturation Spectrum. Neuroimage 2017,DOI10.1016/j.neuroimage.2017.04.045.

(13) Kleesiek J, Urban G, Hubert A, Schwarz D, Maier-Hein K, Bendszus M, Biller A. Deep MRI brain extraction: A 3D convolutional neural network for skull stripping. NeuroImage 2016;129:460–469.

(14) „Deep Convolutional Neural Network for Inverse Problems in Imaging - IEEE Journals & Magazine",can be found underhttps://ieeexplore.ieee.org/abstract/document/7949028/,o. J.

(15) Zhu B, Liu JZ, Cauley SF, Rosen BR, Rosen MS. Image Reconstruction by Domain-Transform Manifold Learning. Nature 2018;555:487–492.

(16) Shajan G, Kozlov M, Hoffmann J, Turner R, Scheffler K, Pohmann R. A 16-Channel Dual-Row Transmit Array in Combination with a 31-Element Receive Array for Human Brain Imaging at 9.4 T. Magnetic Resonance in Medicine 2014;71:870–879.

(17) Zaiss M, Ehses P, Scheffler K. Snapshot-CEST: Optimizing Spiral-Centric Reordered Gradient Echo Acquisition for Fast and Robust 3D CEST MRI at 9.4T. NBM 2017,DOI10.1002/nbm.3879.

(18) Schuenke P, Windschuh J, Roeloffs V, Ladd ME, Bachert P, Zaiss M. Simultaneous Mapping of Water Shift and B1 (WASABI)-Application to Field-Inhomogeneity Correction of CEST MRI Data. Magnetic Resonance in Medicine 2017;77:571–580.

(19) Cox RW, Hyde JS. Software Tools for Analysis and Visualization of FMRI Data. NMR in biomedicine 1997;10:171–178.

(20) Windschuh J, Zaiss M, Meissner J-E, Paech D, Radbruch A, Ladd ME, Bachert P. Correction of B1-Inhomogeneities for Relaxation-Compensated CEST Imaging at 7 T. NMR in biomedicine 2015;28:529–537.
(21) Zaiss M, Windschuh J, Paech D, Meissner J-E, Burth S, Schmitt B, Kickingereder P, Wiestler B, Wick W, Bendszus M, Schlemmer H-P, Ladd ME, Bachert P, Radbruch A. Relaxation-Compensated CEST-MRI of the Human Brain at 7T: Unbiased Insight into NOE and Amide Signal Changes in Human Glioblastoma. Neuroimage 2015;112:180–188.
(22) Møller MF. A scaled conjugate gradient algorithm for fast supervised learning. Neural Networks 1993;6:525–533.
(23) Zaiss M, Schuppert M, Deshmane A, Herz K, Ehses P, Füllbier L, Lindig T, Bender B, Ernemann U, Scheffler K. Chemical Exchange Saturation Transfer MRI Contrast in the Human Brain at 9.4 T. Neuroimage 2018;179:144–155.
(24) Xu X, Yadav NN, Knutsson L, Hua J, Kalyani R, Hall E, Laterra J, Blakeley J, Strowd R, Pomper M, Barker P, Chan K, Liu G, McMahon MT, Stevens RD, van Zijl PC. Dynamic Glucose-Enhanced (DGE) MRI: Translation to Human Scanning and First Results in Glioma Patients. Tomography 2015;1:105–114.
(25) Schuenke P, Koehler C, Korzowski A, Windschuh J, Bachert P, Ladd ME, Mundiyanapurath S, Paech D, Bickelhaupt S, Bonekamp D, Schlemmer H-P, Radbruch A, Zaiss M. Adiabatically Prepared Spin-Lock Approach for T1ρ-Based Dynamic Glucose Enhanced MRI at Ultrahigh Fields. Magnetic Resonance in Medicine 2017;78:215–225.

**Additional Information:**

DeepCEST: 9.4 T Chemical Exchange Saturation Transfer MRI contrast predicted from 3T data - a proof of concept study

Moritz Zaiss[1], Anagha Deshmane[1], Mark Schuppert[1], Kai Herz[1], Philipp Ehses[2], Tobias Lindig[4], Benjamin Bender[4], Ulrike Ernemann[4], Klaus Scheffler[1,5]

**Author contributions statements**

MZ and AD wrote the main manuscript text and MZ prepared figures. PE, MZ created the 9.4T sequence. MZ, KH, and MS optimized and measured the 9.4T data. MZ, KH, and AD optimized and measured the 3T data. MZ optimized and applied the neural networks. TL, BB, and UE were responsible for clinical data, patient recruitment and interpretation of findings. All authors reviewed the manuscript.

**Competing interests**
The authors declare no competing interests.